\documentclass[manuscript,nonacm]{acmart}
\AtBeginDocument{%
  }

\setcopyright{acmlicensed}
\copyrightyear{2026}
\acmYear{2026}

\acmConference[HCXAI@CHI '26]
{Human-centered Explainable AI Workshop (HCXAI), Re-examining XAI in the Era of Agentic AI}
{2026}

\begin{document}

\title{Less Interaction But More Explanation: A Communication Perspective on Agentic AI Interfaces}

\author{Eunchae Jang}
\email{ezj5160@psu.edu}
\orcid{0000-0001-7547-9050}
\affiliation{%
  \institution{Media Effects Research Laboratory, Pennsylvania State University}
  \city{University Park}
  \country{Pennsylvania, USA}
}

\author{S. Shyam Sundar}
\email{sss12@psu.edu}
\orcid{0000-0002-5779-8864}
\affiliation{%
  \institution{Center for Socially Responsible AI, Pennsylvania State University}
\city{University Park}
 \country{Pennsylvania, USA}
}

\renewcommand{\shortauthors}{Jang \& Sundar}

\begin{abstract}
AI systems have long been expected to interact with users—answering questions, generating content, and continuing (social) conversations. Agentic AI, however, breaks from this expectation, as its primary objective is workflow execution on behalf of the users. If a system becomes more agentic, do users need less interaction with the system? Our answer is: less routine back-and-forth, but more communication for oversight and explanation, as agentic AI proactively acts, not just responds. Grounded in a communication perspective, we discuss how users perceive the communicative roles of AI systems (whether as the source of actions or merely a channel), and how this can shape trust. Because agentic AI can play multiple communicative roles, it can complicate this source perception and introduce potential risks. To address this, we propose three types of explanations that agentic AI needs to incorporate (action-process, uncertainty, and coordination), and suggest that customization affordances that allow users to decide when and which explanations they see may be key to preserving human agency as AI autonomy increases.
\end{abstract}



\maketitle

\section{Introduction}
Traditionally, users have developed an expectation of AI as an interactive, communicative system. Generative AI reinforced this expectation and gained widespread adoption in part as they foregrounded interaction, such as back-and-forth conversation and other conversational affordances that keep users "in the loop" \cite{jung2024we, sun2024generative}. Recently, agentic AI has fundamentally shifted human-AI interaction from a primarily conversational partner to an autonomous agent capable of pursuing complex, multi-step goals through iterative feedback loops; adapting to dynamic environments; exercising greater autonomy; and, in some implementations, coordinating work through multi-agent orchestration \cite{acharya2025agentic, hosseini2025role, sapkota2025ai}. Although agentic AI is often built on large language models (LLMs) that enable natural language interaction \cite{sapkota2025ai}, its primary value is not conversation \textit{per se}. Rather, it is oriented toward workflow execution, including clarifying users’ goals and preferences \cite{acharya2025agentic}. In this sense, agentic AI aims to reduce human micromanagement (and thus requires less back-and-forth interaction) by acting on users' behalf. That is, the less a user needs to instruct each step and intervene in execution, the more "agentic" the system is considered to be. This shift raises important questions: if a system becomes more agentic, do users need less interaction with the system? Does increasing AI autonomy inevitably reduce conversation needs because the system should know the user well enough to complete tasks independently? 

When agentic AI reduces the need (or opportunity) for users to communicate, because it takes actions rather than generating responses or content, risks may increase. These include goal misalignment (how the system interprets the user’s intent), errors (mistakes in early steps can compound across a pipeline), and over-trust (users may perceive outcomes as more reliable than warranted). Consequently, agentic AI can inadvertently reinforce a classic “black box” problem. And the stakes are higher because the explanation needs are no longer limited to how an output was generated but extends to what (and how) actions were taken to achieve a goal. Consequently, users may need a clearer explanation to understand the system's decisions while maintaining human autonomy. 

In this position paper, we argue that greater AI autonomy should not reduce communication needs; instead, it demands a rethinking of how agentic AI communicates with users to support accountability and explainability. Thus, we examine agentic AI through a communication perspective, focusing on how users understand the communicative roles of agentic AI, how such perceptions can inadvertently introduce potential risks, and what types of explanations are needed to mitigate these emerging risks.

\section{Understanding Agentic AI's Communication Roles}
To answer the questions posed at the outset of our position paper, we first need a theoretical framework that can explain what communicative roles users may attribute to agentic AI, i.e., how users may perceive agentic AI as the source. This is because communication outcomes, such as trust, user experience, and the need for explanation depend on how AI is perceived as a source. This “source orientation” \cite{sundar2000source} typically lies in whether AI is perceived as a \textit{source} of interaction, or as merely a \textit{medium or channel} for interactions between humans \cite{sundar2022rethinking}. 

Sundar and Lee’s 4C typology \cite{sundar2022rethinking}–classification of AI involvement in human communication–offers a useful conceptual lens to understand this. The 4C typology is defined by two axes: (a) the communication context (mass vs. interpersonal) and (b) AI’s role in the communicative process (communicator vs. mediator), creating four roles: \underline{AI Creator} (mass × communicator; AI produces messages for broad audiences, such as virtual influencers or robot reporters), \underline{AI Converser} (interpersonal × communicator; AI engages users in dialogue, such as chatbots or smart speakers), \underline{AI Curator} (mass × mediator; AI filters, ranks, recommends, or moderates content, such as recommender systems or content moderation tools), and \underline{AI Co-Author} (interpersonal × mediator; AI assists humans in creating messages, such as drafting or suggesting language via auto-correction or generative AI writing tools). In the case of agentic AI, it expands the typology by traversing multiple roles across a workflow, and it can also collapse the boundaries of roles, introducing potential risks around source perceptions.

\textbf{Expansion}. Imagine an agentic AI tasked with managing a company's external communications strategy, such as "handle our product launch communications" as a goal. At the outset, the system may function as an \textit{AI Converser}, engaging the communications manager to clarify the launch timeline, target audiences, key messages, and message formats. As it monitors the environments, from tracking media coverage for competitors, analyzing target audiences to flagging trending narratives and formats, it operates as an \textit{AI Curator}, filtering and ranking information from broad information sources to inform strategic decisions. When it drafts communication pitches for the communications manager to send to specific journalists or stakeholders, it serves the role of \textit{AI Co-Author}, assisting in producing messages where the human remains the apparent source. And when it generates and publishes press releases, social media posts, or blog content for broad public audiences, it fulfills the role of \textit{AI Creator}. 

\textbf{Collapse}. Agentic AI can also collapse these distinctions in ways that prior AI systems have not. When a system autonomously decides what to do, when to act, and how to proceed across a multi-step pipeline without human intervention at each step, it simultaneously executes human goals, while playing multiple roles on its own. Here, the question of whether AI is the source or channel becomes difficult to answer, even for attentive users. We discuss further how this ambiguity of source orientation can result in potential risks, in the following section.

\section{Source Orientation of Agentic AI and Potential Risks}
When users cannot reliably identify the AI system as the originating source of action and perceive it as carrying out their goals, it can have significant implications for information processing \cite{sundar2022rethinking, dehnert2022persuasion}. For example, users are less likely to scrutinize its actions, question accuracy of its decisions and outcomes, or feel the need for explanations. As such, the misattributed source orientation can result in several risks. First, it creates \textbf{attribution problems}: when users are not familiar with the agentic AI architecture, they may be unsure or unaware of whether an outcome reflects the higher-level agent’s decisions (i.e., a single agent's decision), the sub-agents' decisions (i.e., multiple agents involved), or is largely influenced by external tools and data sources. This makes it difficult for users to trace the system's actions, decisions, and  outcomes when needed. This issue can further produce \textbf{accountability gaps}. If users perceive agentic AI as a single source while the system is in fact coordinating multi-step processes (with multiple agents), they may underestimate the uncertainty or disagreement between agents underlying the result. At the same time, when multiple specialized sub-agents are identified as being involved, it can result in a different set of psychological vulnerabilities. For example, users may treat internal agreement among agents as a credibility signal. In such cases, multi-agent coordination can inadvertently function as a cue, where it triggers the \textit{consensus heuristic} ("since multiple agents worked and reached the output, it must be credible'') \cite{chen1999motivated}. When sub-agents are framed as specialists, it can trigger the \textit{authority heuristic} ("since a specialized sub-agent worked on each step, it must be credible") \cite{sundar2008main}. These cognitive heuristics (mental shortcut) may lead to \textbf{over-trust} that exceeds warranted system reliability. Liao and Sundar \cite{liao2022designing} further outlined heuristics that can lead to over-trust, such as the \textit{confirmation heuristic} (accepting outputs when they are aligned with users' worldviews), highlighting the need to identify or develop warranted trustworthiness cues that support appropriate trust calibration.

\setlength\parindent{12pt} Taken together, it suggests that how users perceive agentic AI can bring about different risks. If users perceive agentic AI as a single source despite its operation as a multi-agent system, the system needs mechanisms that clearly show the end-to-end workflow and the actions taken. If users perceive it as multi-agent sources, or are confused about the true source of actions or outputs, the system should clarify the relevant provenance and decision points to support informed oversight and, when appropriate, user intervention. To address these concerns, we propose three types of explanations, along with when users might need each of them (before, during, and after task execution), in the next section.

\section{Explanation Needs in Agentic AI From a Communication Perspective}
Given the considerable risk of source misattribution in agentic AI, due to its characteristic of playing multiple communicative roles (Creator/Converser/Curator/Co-author), explanations should make (1) agents' role assignments and transitions, and the provenance of actions taken in each role, visible across the workflow, (2) uncertainty that can arise from each communicative role clear, and (3) multi agent-based decision-making more explicit so that "coordination" is not mistaken for validation.

\begin{itemize}
\item \textbf{Action-Process Explanations} address the attribution problem by making visible what the system did, in what order, why, and which communicative role (4C) it enacted at each step. The focus is the step-by-step trace: how the system interpreted the user’s goal, what plan it generated, which steps it executed, what tools/APIs were used, and what intermediate artifacts were produced. Importantly, in a multi-agent system, these explanations should label which sub-agent performed each step and tag that step with the relevant 4C role (e.g., “Curator: retrieved competitor coverage”; “Co-Author: drafted journalist pitch”; “Creator: generated final press release”). By disclosing the communicative role of agentic AI throughout the process, action-process explanations help users answer the attribution question―which roles, and which aspects within those roles―may warrant extra attention and scrutiny. For example, a sub-agent organizing information from online sources, labeled as "Curator," may make users want to check source credibility for accuracy, while another sub-agent labeled as "Co-Author" may prompt users to verify whether the system's assistance in writing aligns with their initial intention. Indeed, current ChatGPT's \textit{Agent mode} (which utilizes the agentic AI system) provides a visualized (real-time, micro) step-by-step explanation of the actions that the model has taken; however, considering the longer time to provide the final output, the steps shown are too many for users to review, ironically burdening them and potentially reducing their intention for oversight. If presented with visualization \cite{chatti2024visualization} and role labels at each step, such explanations can reduce cognitive overload and make role shifts legible, supporting meaningful human autonomy without requiring micromanagement at every step. 

\setlength\parindent{12pt} More importantly, users can benefit more when (planned) action-process explanations are provided "before” execution \cite{chen2024explain}, so that they can clarify assumptions and ensure whether their intents are correctly interpreted in advance, particularly when poor execution is costly. This also means that with agentic AI, more upfront back-and-forth interaction may be needed. Indeed, many agentic AI systems do not require users to specify their goals after initial prompt, even if it lacks details of the goals, which will lead to poor execution. Additionally, a recent study \cite{lee2025while} revealed that users tend to attribute different mental models while waiting for LLMs to generate outputs, such as one understands it as "\textit{analyzing} what I typed into it," while the other perceives it as "trying to \textit{search} for results." This suggests that making the system’s (planned) steps and actions clear prior to execution (perhaps through more back-and-forth interaction) can help users form and apply more accurate mental models and better know when to intervene.

\item \textbf{Uncertainty explanations} address the accountability gap by communicating the degree of confidence, assumptions, uncertainties, and/or limitations present in the system's decision-making process/outcome \cite{phillips2021four}. Because agentic AI often operates across multi-step pipelines, uncertainty can accumulate at each decision point. For example, plausible but unverified assumptions can propagate into later steps. Uncertainty explanations should therefore target what the system does not know, what it assumes, what source(s) it relies on, and where its output is most likely to be unreliable. Acknowledging that the model may not know uncertainties it may encounter while executing, presenting uncertainty per 4C agents' role (e.g., "Curator agent relied on limited sources" or "Creator agent extrapolated the findings from studies with small samples") during or after execution may help users avoid over-trusting the final outcome. 

\setlength\parindent{12pt} In addition, when delivering those uncertainties to users, they need to be given control to intervene during the execution when needed. Many agentic AI systems, unfortunately, do not support such options. For example, ChatGPT's Agent mode provides two options: the "pause" button to cease the execution, or the "take over browser" button, in which users need to control/execute by themselves. This lack of interaction affordance to instruct the model during the process may limit its successful execution. Uncertainty explanations can also be useful after execution. By showing which part of the workflow/or decision may be most susceptible, users are better able to decide where to focus their oversight.

\item \textbf{Coordination explanations} address potential over-trust risks by clarifying how multi-agent outputs are coordinated. The focus here is not the step-by-step trace, but how decisions are made or will be made: how the system decomposes the task across sub-agents, whether sub-agents are truly independent or reliant on external tools/sources, and how disagreements between agents are resolved. This is the explanation type that can directly counter the “consensus heuristic” and “authority heuristic,” described on p. 3, by making clear whether agreement reflects true consensus among agents, or correlated model assumptions, which may need more warranting. To support accurate source orientation, coordination explanations can explicitly describe how the agents interacted (e.g., whether Curator agents provided evidence to a Co-Author agent, or whether a Creator agent could publish without human approval). Ideally, the coordination explanation can be provided prior to execution as well, offering user control to assign and specify the roles for agents and add/remove certain agents' roles (e.g., adding an agent that specifically accounts for verifying the accuracy of information using XYZ tools). 
\end{itemize}

Importantly, these three types of explanation may not always be needed simultaneously. While a single type may fail to satisfy different explanation needs of users \cite{arya2019one}, over-explanation can also undermine oversight by overwhelming users with information overload \cite{KallinaSingh2025ResponsibleXAI}. Furthermore, poorly designed explanations can themselves become a source of miscalibration, primarily resulting in over-trust \cite{naiseh2023different}. And this over-trust issue is especially consequential as the AI systems are biased, and research has noted that lay users fail to notice a visibly skewed bias, even when provided with clear visual explanation \cite{chen2025racial}, and the biases in the AI models are persistent \cite{guo2025exposing}. Ultimately, the appropriate level, timing, and types of explanation may depend on the types of task, the stakes of the task, and the degree of autonomy exercised before/during execution \cite{chen2024explain, phillips2021four, KallinaSingh2025ResponsibleXAI}.

As a solution, previous studies \cite{lee2025shorter, liao2022designing} have shown that affording customization, in this case the type of explanations users want to see and/or when to intervene, can increase users agency (control), which is essential particularly when AI exerts more autonomy than human. For example, Lee and colleagues \cite{lee2025shorter} showed that customizing the tone and length of generative AI's answers increased perceived user control, compared to the absence of such customization, showing greater user satisfaction and source credibility of AI. Last but not least, we need more user studies that investigate how agentic AI may shape user trust, what kinds of explanations best protect users from potential risks, and how agentic AI can be designed to preserve human autonomy in support of XAI.

\section{Conclusion}
Agentic AI shifts human-AI interaction from primarily conversational exchange to goal-directed workflows, reducing the need for constant back-and-fourth communication. We argue that the capability of agentic AI to perform multiple communicative roles within a single task can introduce risks such as attribution problems, accountability gaps, and over-trust. To mitigate these risks and support calibrated trust, we propose three types of explanations that agentic AI systems should offer to users: action-process, uncertainty, and coordination explanations. Finally, as a design approach for supporting users’ explanation needs, we suggest allowing users to customize both the type and timing of explanations they want to see, thereby preserving their agency.

\bibliographystyle{ACM-Reference-Format}
\bibliography{reference}

\end{document}